# Leveraging Low-Energy Structural Thermodynamics in Halide Perovskites

Bryan A. Rosales,[1] Kelly Schutt,[1] Joseph J. Berry,[1] and Lance M. Wheeler[*,1]

[1]National Renewable Energy Laboratory, 15013 Denver West Parkway, Golden, CO 80401, USA.
AUTHOR INFORMATION

**Corresponding Author**

*E-mail: Lance.Wheeler@nrel.gov

**ABSTRACT**

Metal halide perovskites (MHPs) combine extraordinary optoelectronic properties with chemical and mechanical properties not found in their semiconductor counterparts. For instance, they exhibit optoelectronic properties on par with single-crystalline gallium arsenide yet exhibit near-zero formation energies. The small lattice energy of MHPs means they undergo a rich diversity of polymorphism near standard conditions similar to organic materials. MHPs also demonstrate ionic transport as high as state-of-the-art battery electrodes. The most widespread applications for metal halide perovskites (e.g. photovoltaics and solid-state lighting) typically view low formation energies, polymorphism, and high ion transport as a nuisance that should be eliminated. Here, we put these properties into perspective by comparing them to other technologically relevant semiconductors in order to highlight how unique this combination of properties is for semiconductors and to illustrate ways to leverage these properties in emerging applications.

**TOC GRAPHICS**

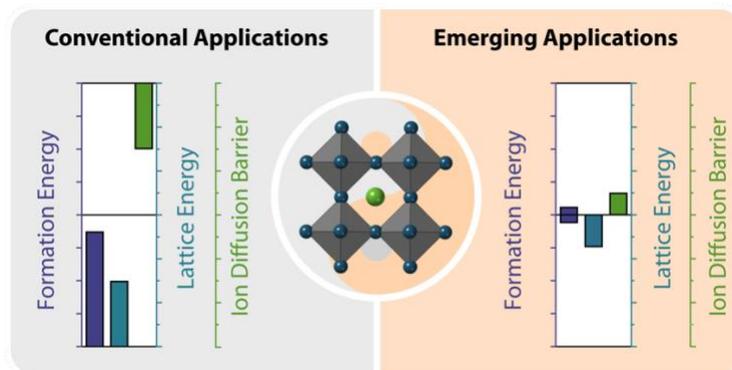

Metal halide perovskites (MHPs) exhibit outstanding optoelectronic properties[1-6] featuring long carrier lifetimes and diffusion lengths, tunable bandgaps, and exceptional defect tolerance that result from a crystalline-like electronic band structure combined with liquid-like ionic lattices.[7] The prototypical MHP structure is represented by $AMX_3$ (A = methylammonium ($MA^+$), formamidinium ($FA^+$), $Cs^+$, etc.; M = $Pb^{2+}$, $Sn^{2+}$; X = $I^-$, $Br^-$, $Cl^-$), characterized by a 3D network of interconnected $[MX_6]^{4-}$ octahedra with monovalent A-site cations occupying interstitial sites (Figure 1). The connectivity of the $[MX_6]^{4-}$ octahedra dictates the observed optoelectronic properties and can range from the fully connected perovskite phase (3D) to partially connected (2D/1D) and fully isolated (0D) non-perovskite phases.[8] The MHP lattice also accommodates a variety of dopants that influence the optoelectronic properties by disrupting the $[MX_6]^{4-}$ octahedral network and by interacting with the soft, polarizable lattice.[9]

MHPs undergo a variety of structural changes near standard optoelectronic device operational conditions. Though MHP-based photovoltaic (PV) devices have achieved a remarkable >25% power conversion efficiency (PCE) within a decade of research,[10] fundamental material properties are easily changed by exposure to humidity, heat, electric fields, and light, which, if not managed, result in undesirable transitions between crystal structures, changes in electronic structure, ion migration, and irreversible decomposition.[11-13] Transformation of material properties is detrimental to many established applications of MHPs spanning PVs, electronics, light-emitting diodes (LEDs), lasers, and thermoelectrics (Figure 1). Considerable effort has gone into improving and managing transformations in MHPs to prevent changes in crystalline or electronic structure through formation energy manipulation,[14-20] reducing defect density,[21] and incorporating hydrophobic organic cations.[22]

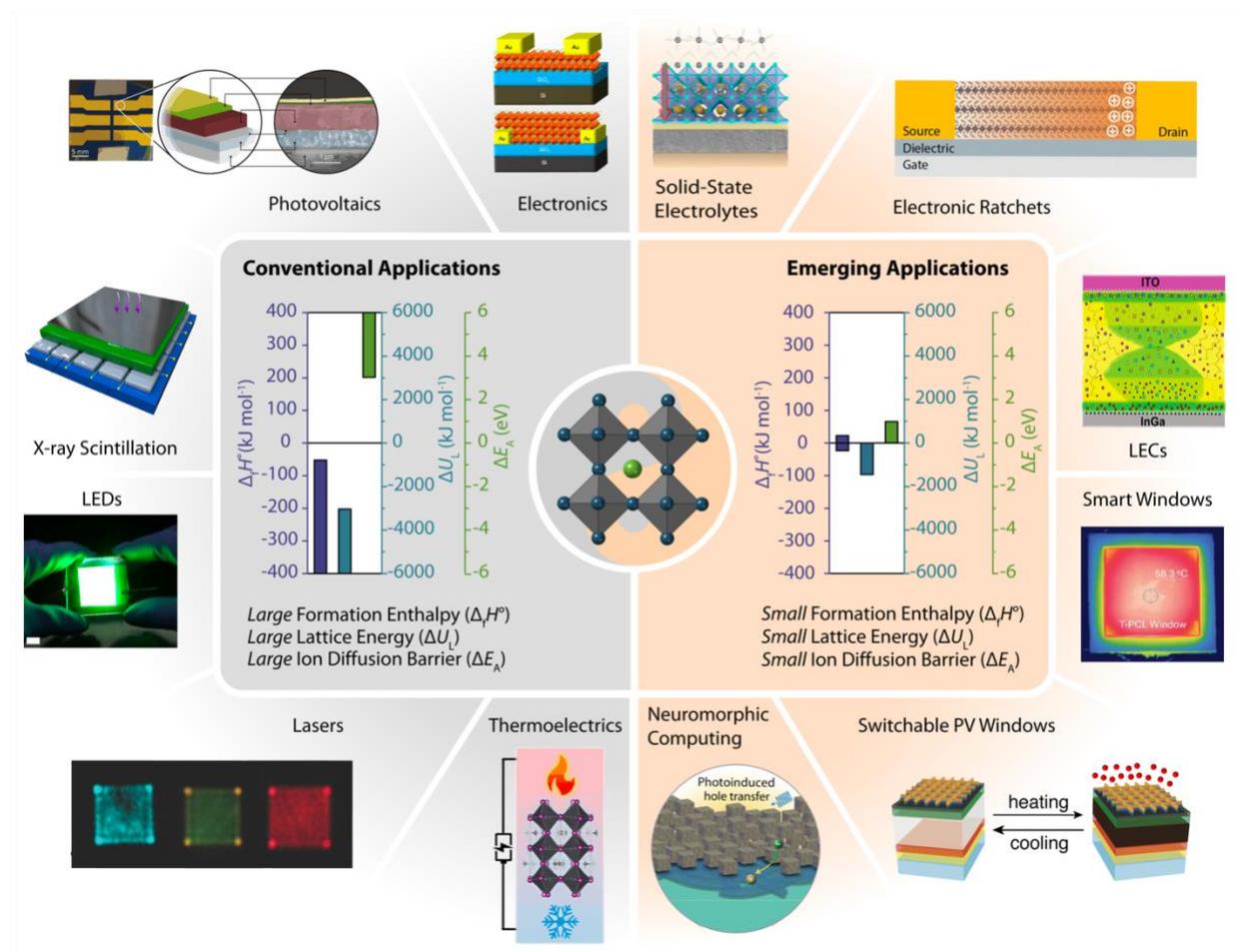

**Figure 1.** Schematic overview comparing conventional MHP applications that require high resilience to material property transformations with emerging MHP applications that take advantage of material property transformations. Inset images were reprinted with permission from refs. 23-29 copyright Springer Nature, refs. 30-32 copyright Wiley-VCH, ref. 33 copyright American Chemical Society, and ref. 34 copyright American Association for the Advancement of Science.

In MHP literature, "stability" generally refers to resilience to changes in material properties. In this perspective, we discuss the fundamental properties of MHP materials that underpin stability by comparing the formation energy, lattice energy, and ionic transport of MHP materials to technologically relevant semiconductors. This comparison illuminates just how unique MHP properties are across materials science. Instead of the stabilization-centric narrative driven by PV literature, we highlight MHPs as switchable and stimuli-responsive semiconductors[35] that enable new applications spanning switchable PV smart windows,[28, 29] electronic ratchets,[31] solid-state electrolytes,[27] light-emitting electrochemical cells (LECs),[32] and artificial synapses for neuromorphic computing[34] (Figure 1).

*Formation Energy of MHPs.*

Unlike traditional semiconductors, MHPs are entirely solution processable at low temperatures and atmospheric pressure,[13,36] and do not require high-temperature melt-processing, vapor phase growth in a UHV chamber, or forming inks from nanocrystals that were pre-synthesized at higher temperatures. MHP single crystals can also be synthesized at room temperature,[37] and engineering the acidity of the growth solution enables low temperature growth with reduced defect densities.[38] In contrast to MHPs, conventional semiconductors such as Si are grown by heating above their melting point of 1414 °C while III-V semiconductors require vapor phase growth above 550 °C in ultrahigh vacuum.[39] Low energy intensity processing promises reduced fabrication costs and increased manufacturing throughput for an array of applications.

Low-energy synthesis is a consequence of low formation energies in MHPs. The Gibbs free energy of formation ($\Delta_f G°$) is a measure of the thermodynamic driving force for synthesis under standard conditions (25 °C and 1 atm = 101,325 Pa) without external stimuli (e.g., light, heat, $H_2O$, $O_2$, etc.). $\Delta_f G°$ is represented by:

$$\Delta_f G° = \Delta_f H° - T\Delta_f S° \qquad (1)$$

where enthalpy of formation ($\Delta_f H°$) is the energy released or consumed when one mole of a product is synthesized from reactants and can be thought of as a measure of the strength of bonds broken and formed, while entropy of formation ($\Delta_f S°$) describes the degree of compositional and energetic disorder in the system. Formation of a compound is thermodynamically favorable for $\Delta_f G° < 0$, and unfavorable for $\Delta_f G° > 0$. We compare the formation energy of MHPs to binary and elemental semiconductors due to their similar applications, with ternary oxide perovskites as a direct structural comparison to MHPs (Figure 2a). MHPs have slightly negative $\Delta_f G°$ values in the range of 0 to -15 kJ mol$^{-1}$, whereas conventional semiconductors exhibit a large driving force relative to precursors, with deeply negative $\Delta_fG°$ values from -50 to -1500 kJ mol$^{-1}$ (Figure 2b).

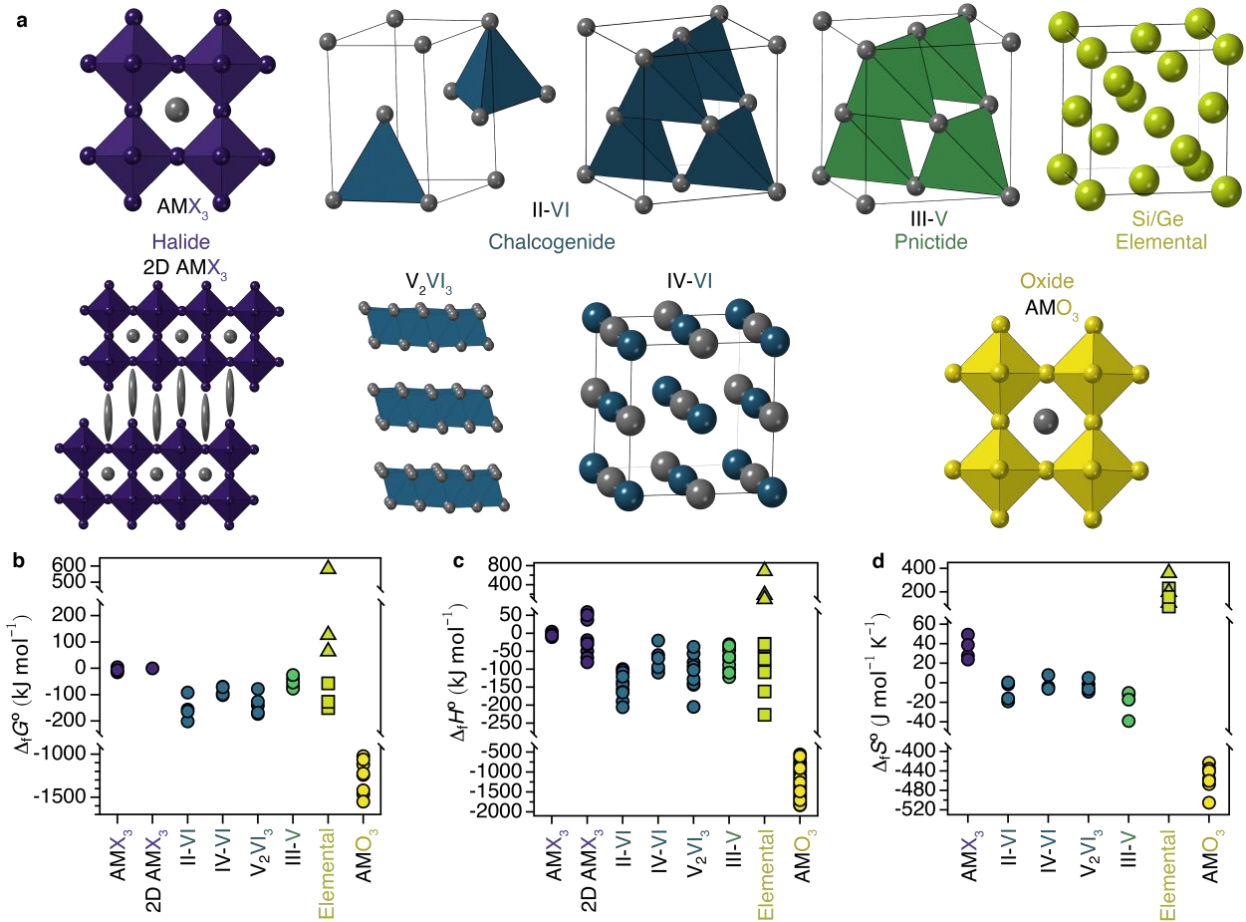

**Figure 2.** (a) Unit cell comparison of the halide, chalcogenide, pnictide, elemental, and oxide semiconductors compared in this study. Comparison of (b) Gibbs free energy of formation ($\Delta_fG°$), (c) enthalpy of formation ($\Delta_fH°$), and (d) entropy of formation ($\Delta_fS°$) values of MHPs to those of other conventional semiconductors. Values for elemental semiconductors were obtained from oxide precursors (triangles) or from chemical vapor deposition (CVD) precursors (squares). Values for chalcogenide, pnictide, and oxide semiconductors were obtained from elemental precursors.

The formation energy of inorganic compounds is often discussed in terms of $\Delta_fH°$ due to the inherently strong bonds created when these compounds are formed. Elemental semiconductors prepared by chemical vapor deposition (CVD, squares) as well as binary chalcogenide (II-VI, IV-VI, V$_2$VI$_3$) and pnictide (III-V) semiconductors exhibit $\Delta_fH°$ values in the range of -50 to -300 kJ mol$^{-1}$, whereas ternary oxide perovskites (AMO$_3$) are as negative as -2000 kJ mol$^{-1}$ (Figure 2c).

Large, negative $\Delta_fH°$ values suggest these compounds are unlikely to decompose into precursors due to their strong bonding. Indeed, these compounds are highly stable at > 500-1000 °C, leading to their commercial use in challenging environments.[40, 41]

MHPs are typically synthesized from the reaction between alkylammonium or metal halide salts with metal dihalide salts, as described by the following equation:

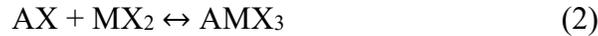

$$AX + MX_2 \leftrightarrow AMX_3 \qquad (2)$$

In contrast to conventional semiconductors, the $\Delta_fH°$ term of MHPs is between +5 and -10 kJ mol$^{-1}$ (Figure 2b), with near-equal driving forces between formation and dissociation into precursors. Resiliency of MHPs to decomposition follows the intuitive trend Cl > Br > I, in agreement with the Pb-halide bond strengths and redox potentials.[42-44] Whereas MHPs are produced from precursors with similar ionic bonds, the precursors for chalcogenide, pnictide, and oxide semiconductors are elemental, metal salts, or organometallics with products containing bonds to metal centers with higher valency anions, resulting in stronger bonding environments.

The $\Delta_fH°$ term typically dominates the $\Delta_fS°$ term in technologically relevant semiconductors due to large, negative $\Delta_fH°$ values and small $\Delta_fS°$ values (0 to -20 J mol$^{-1}$ K$^{-1}$) which are often ignored (Figure 2d). In contrast, the $\Delta_fS°$ term of MHPs cannot be ignored. MHPs exhibit positive $\Delta_fS°$ (+20 to +60 J mol$^{-1}$ K$^{-1}$) values that introduce a greater driving force than the near-zero $\Delta_fH°$.[42, 45] Entropy is a facile method to improve the operational stability of MHPs in solar cells because $\Delta_fS°$ can be easily increased through alloying and/or doping multiple cations and halides.[18] Despite high susceptibility to decomposition, MHP-based PV devices have maintained > 85% of their initial performance after 1000 h[46] of illumination, with the current record at 10,000 h (> 1 yr).[47]

The low formation energy of MHPs means small disturbances can lead to decomposition of the MHP into its precursors or transformation into non-perovskite phases. The most widely investigated transformations in MHPs involve H$_2$O.[48-55] Reversible hydration of MHPs under controlled humidity occurs according to the following reaction:

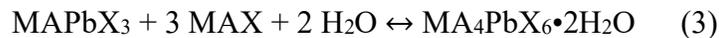

$$MAPbX_3 + 3\ MAX + 2\ H_2O \leftrightarrow MA_4PbX_6 \cdot 2H_2O \qquad (3)$$

3D MAPbI$_3$ films prepared with excess MAI ($E_G$ = 1.80 eV) form the 0D hydrated phase MA$_4$PbI$_6$•2H$_2$O ($\lambda_{max}$ = 370 nm = 3.35 eV) above 40% RH and MAPbI$_3$ can be regenerated by dehydrating above 75 °C (Figure 3a).[51] Hydration also causes decomposition of MAX by pushing equilibrium towards volatile MA + HX, forming solid PbI$_2$ upon leaving the film.[12]

While detrimental to PV, reversible hydration has been leveraged for applications that benefit from stimuli-responsiveness. For example, switchable PV devices can generate electricity in the colored dehydrated state and offer high light transmittance in a hydrated state.[49-50] In contrast to hydrate phase formation in MA-based MHPs, H$_2$O triggers reversible dimensionality tuning of Ruddlseden-Popper phases in FA-based MHPs. Whereas most chromic films offer a single-color transition, this mechanism enables switching between multiple colors including yellow, orange, red, brown, and white/colorless by turning the thickness of the layers.[55]

Intercalating species that H-bond with the MHP lattice, including methanol (MeOH) and amines, can also drive switchable devices.[29, 56, 58-63] MeOH exhibits weaker H-bonding with the lattice relative to H$_2$O, reducing switching temperature from 70 °C to 50 °C. Thermochromic windows utilizing methanolation and PAA have visible light transmittance (VT) of 31% in the colored state and 84% in the bleached state, with ~80% of the initial VT after 200 cycles.[57] Amine vapor transforms the 3D MHP into a lower dimensional MHP through complex formation. This process is reversed at room temperature upon removing the amine vapor (Figure 3a). Sealing the MAPbI3•xCH3NH2 complex to prevent amine escape enables thermochromic windows that darken upon complex dissociation under solar heating above 60 °C, with a champion PCE of 11.3% and a VT ranging from 3 to 68%.[29]

*Polymorphism in MHPs.*

MHP materials adopt highly symmetric cubic structures at elevated temperatures. Corner-sharing MX$_6$ octahedra tilt with respect to one another at decreasing temperatures, which results in several polymorphic transitions to lower-symmetry structures (Figure 3a). Polymorphic transformations in MHPs include those that maintain the bonding environment of the original structure but with a change in symmetry, and those that change both the bonding environment and symmetry. In general, the closer a polymorph is to the original structure, the more similar the optoelectronic properties will be.[64] For example, α (cubic), β (tetragonal), and γ (orthorhombic) polymorphs (Figure 3a) undergo a symmetry change due to octahedral tilting while maintaining a similar bonding environment, resulting in only a 0.2-0.3 eV bandgap difference.[64] MHPs also exhibit two polymorphs near standard conditions with a significant change in bonding environment (3D ↔ 1D) resulting in large differences in optoelectronic properties: a 3D black cubic phase (α, $E_G$ = 1.45-1.75 eV) and a 1D pale-yellow orthorhombic (Cs-based) or hexagonal (FA-based) phase (δ, $E_G$ = 2.40-2.85 eV) (Figure 3a).

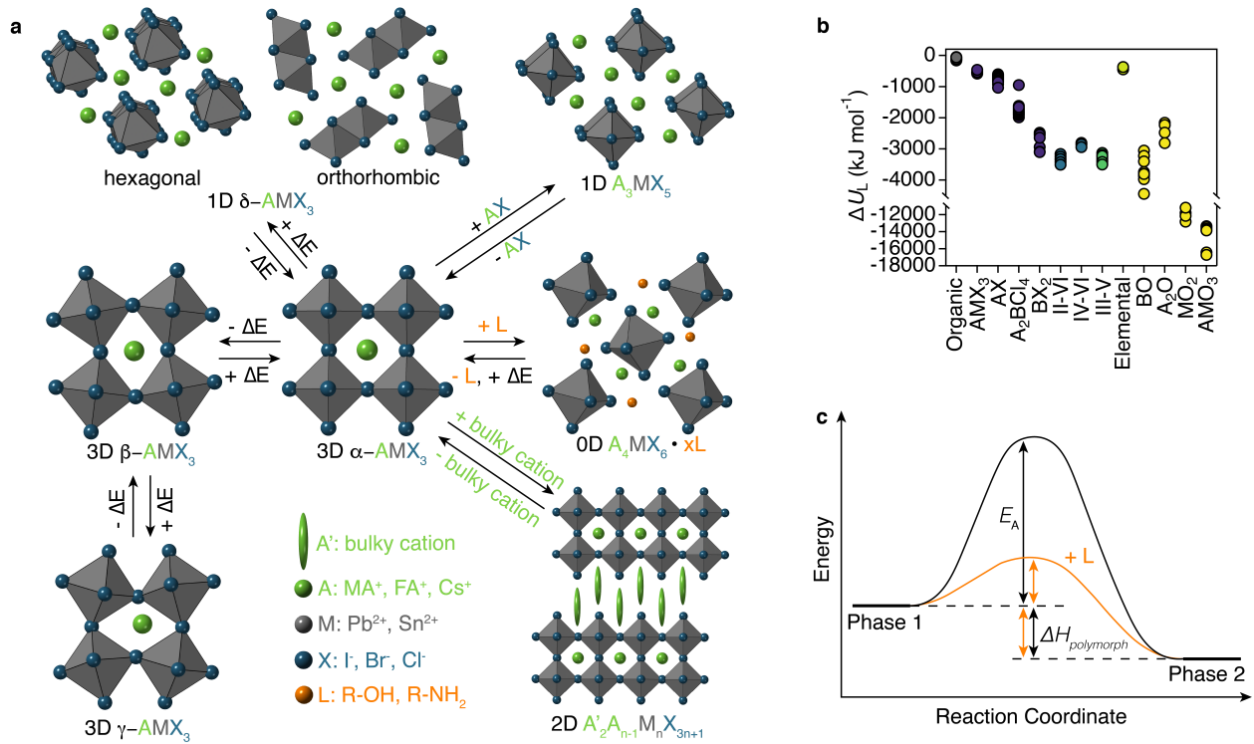

**Figure 3.** (a) Illustration of transformations between the 3D MHP phase with non-perovskite, low-dimensional 2D, 1D, and 0D structures. (b) Comparison of lattice energy values of MHPs to those of other conventional semiconductors. B is a divalent group II or transition metal. (c) Reaction coordinate diagram describing the thermodynamics of phase transformations in MHPs. The activation energy barrier is decreased upon exposure to an H-bonding molecule (L).

The temperature ranges a specific polymorph exists in depends on the composition of the MHP. In general, the temperature range increases from I > Br > Cl, from Pb > Sn, and from FA > Cs > MA.[64] Inorganic solids generally exhibit > 100 °C temperature differences between polymorphs while organic solids exhibit many polymorphs over smaller temperature ranges (~15 °C).[65] MHP exhibit small energy differences between polymorphs (5-14 kJ mol$^{-1}$),[66, 67] comparable to organic materials (< 7 kJ mol$^{-1}$)[68].

Polymorphism reflects lattice energy ($\Delta U_L$), the energy released when infinitely separated ions in the gaseous state under vacuum coalesce to form an ionic lattice. A more negative $\Delta U_L$ indicates stronger cohesive forces whereas a near-zero value of $\Delta U_L$ indicates a weaker lattice and a small change in energy for transformations to occur. $\Delta U_L$ for most ionic semiconductors is significantly larger than the equivalent cohesive energies ($\Delta U_C$) of covalent elemental semiconductors and organic compounds. $\Delta U_L$ of binary chalcogenide, pnictide, and oxide compounds is typically in the range of -3000 to -5000 kJ mol$^{-1}$, and up to -18000 kJ mol$^{-1}$ for oxide compounds (Figure 3b). Metal halides compounds (AX, A$_2$BCl$_4$, BX$_2$) exhibit smaller $\Delta U_L$ values of -500 to -3500 kJ mol$^{-1}$. $\Delta U_L$ of MHPs is significantly lower, -450 to -600 kJ mol$^{-1}$,[69] and more like organic compounds or elemental semiconductors than ionic compounds. These lower values of $\Delta U_L$ show less energy is needed to form polymorphic crystal structures.

Research on conventional applications of MHPs seek to mitigate phase transformations to maintain the desired phase. Though α-phase MAPbI$_3$ launched perovskite PV, FA-rich compositions now yield the highest efficiencies,[70] and Cs-rich compositions are desirable for their thermal stability. α-CsPbI$_3$ and α-FAPbI$_3$ phases are only thermodynamically stable at high temperatures and will spontaneously convert into δ-CsPbI$_3$ and δ-FAPbI$_3$ phases at room temperature on the order of days.

We illustrate polymorphic transformations in MHPs by a reaction coordinate diagram where an activation energy ($E_A$) barrier must be overcome to transform the original structure (phase 1) into a perovskite or non-perovskite structure (phase 2) with an energy difference of $\Delta H_{polymorph}$ (Figure 3c) that is small between different MHP polymorphs. For example, several 2D FA$_2$PbX$_4$ polymorphs exist as chains of corner sharing [PbX$_4$]$^{2-}$ octahedra whose connectivity can vary from (1x1), (3x2) and (3x3) step-like structures in either standard or eclipsed layers with bandgaps only slightly varying within a 0.2 eV range.[71] Because the energy landscape is so small, crystallization of a specific polymorph is often kinetically controlled, rather than thermodynamically, where the less energetic phase is trapped by the energy barrier. For instance, a specific polymorph can be obtained by varying the solvent evaporation rate or by exposure to antisolvent.[72] For energetically unfavorable phases, $E_A$ can be increased, mitigating the transition into the favored phase.

$\Delta H_{polymorph}$ can be manipulated by controlling composition or crystal size. For example, crystal symmetry breaking in $CsPb(I_{1-x}Br_x)_3$ polymorphs leads to anisotropy in carrier transport, high power factor, and ultralow thermal conductivity resulting in promising thermoelectric figures of merit (ZT) of up to 1.7 at room temperature.[73] Cation vacancies lead to a decrease in the energy difference between the α- and δ-phases.[74] Surface strain introduced by synthesizing < 30 nm $CsPbX_3$ quantum dots favors the visibly absorbing α-$CsPbI_3$ phase over the wider bandgap δ-$CsPbI_3$ leading to record PV efficiencies and operational stability for inorganic perovskite PV.[75, 76] Tailored surface chemistry has a similar effect to favor formation of 3D β-$CsPbI_3$ for PV applications.[77] Multi-phase films have also found utility with α-$FAPbI_3$ nanocrystals precipitated from δ-$FAPbI_3$ forming a type 1 heterojunction that boosted LED performance by a factor of 58 compared to bulk-$FAPbI_3$ films.[78]

In contrast to efforts to isolate target phases for conventional applications, low-energy polymorphism enables an array of emerging applications. Exposing MHPs to a trigger molecule that H-bonds to the lattice facilitates ionic reorganization by reducing $E_A$,[28] reducing the transition temperature and increasing transformation kinetics. These transformations typically require heat to reverse, suggesting $E_A$ is sufficient to prevent spontaneous conversion back to the original phase.[11, 79] For instance, the α-to-δ transition is reduced from days to hours upon exposure to humidity, which lowers $E_A$ by initiating ionic reorganization and defect formation at the grain boundary-water interface.[28] The α-phase can be regenerated from the δ-phase upon thermal annealing at high temperatures.[11, 79] Thermochromic PV windows that switch between α-$CsPbIBr_2$ (VT = 35.4%, PCE = 4.7%) and δ-$CsPbIBr_2$ (VT = 81.7%, PCE = 0.15%) for over 40 cycles have been fabricated with the α-to-δ transition occurring at 60% RH over 10h and the δ-to-α transition occurring at 150 °C.[28] MHP-based phase change memory (PCM) has been demonstrated by exploiting optoelectronic variability among polymorphs in perovskites exposed to humidity, light, oxygen, and solvents.[80, 81] A 2D MHP exhibited a large shift in the absorption onset between the amorphous ($T_m$ 173 °C) and crystalline phases ($T_c$ 101 °C), which may allow for Joule heating PCM and simple device integration, but the electronic properties accompanying this phase change have not yet been investigated.[82]

*Ion Transport in MHPs.*

Mass transport is undesirable for conventional optoelectronic devices because ion transport competes with carrier transport. In PV devices, mobile ion accumulation impacts carrier extraction at the electrical interfaces by screening or permitting the formation of defects that cause non-radiative recombination. Despite these concerns, PVs based on MHPs exhibit comparable PCE's to state-of-the-art PV materials[10] while exhibiting high levels of solid-state ion transport even at room temperature.[83, 84] Anomalous properties in MHPs as a result of high ion transport include current-voltage hysteresis, above-bandgap photovoltages, light-induced phase segregation, self-healing, and rapid chemical conversion between cations and anions.[84] Reducing defect density or introducing kinetic barriers to ion transport suppresses ion transport in MHPs, with methods including mixed-cation and mixed-halide alloys,[85] incorporating 2D MHP whose interlayer spacing blocks ion transport,[22] post-synthetic treatment with iodine vapor,[86] doping with K or Rb ions,[87, 88] or exposure to elevated pressure.[19]

The energy input required for ions to migrate throughout a lattice is described by the activation energy of diffusion ($E_{A,d}$) through the following equation:

$$D_{ion} = D_0 e^{\left(\frac{-E_{A,d}}{k_B T}\right)} \qquad (5)$$

where $D_{ion}$ is the diffusion coefficient of an ion, $D_0$ is the temperature-independent prefactor, $k_B$ is the Boltzmann constant, and $T$ is the temperature. $E_{A,d}$ can describe both intrinsic diffusion (diffusion of ions that makeup the lattice) and extrinsic diffusion (diffusion of external ions throughout the lattice). Conventional binary and elemental semiconductors typically exhibit intrinsic $E_{A,d}$ values > 1.0 eV (Figure 4a, circles); intrinsic ion diffusion is uncommon in these materials. In contrast, MHPs exhibit intrinsic $E_{A,d}$ values < 1.0 eV for both the A-site and X-site. The X-site exhibits exceptionally low $E_{A,d}$ values ranging from 0.2-0.5 eV, which is on the same order of magnitude as Li$^+$ diffusion in state-of-the-art Li-ion battery anode (C$_x$Li) and cathode (Li$_{1-x}$ClO$_2$, Li$_{1-x}$Mn$_2$O$_4$, Li$_{1-x}$FePO$_4$) electrodes (Figure 4a). These low $E_{A,d}$ values confer remarkable properties such as complete A- and X-site exchange on the order of seconds. The low $E_{A,d}$ values also lead to phase segregation under illumination for mixed MHP compositions, with a driving force provided by the free energy reduction for photocarriers trapped in lower bandgap phases.[19, 89] In contrast, the M-site $E_{A,d}$ values are > 2.0 eV and therefore remain stationary. MHPs also undergo extrinsic ion diffusion with Li$^+$ exhibiting $E_{A,d}$ values of ~0.1 eV.

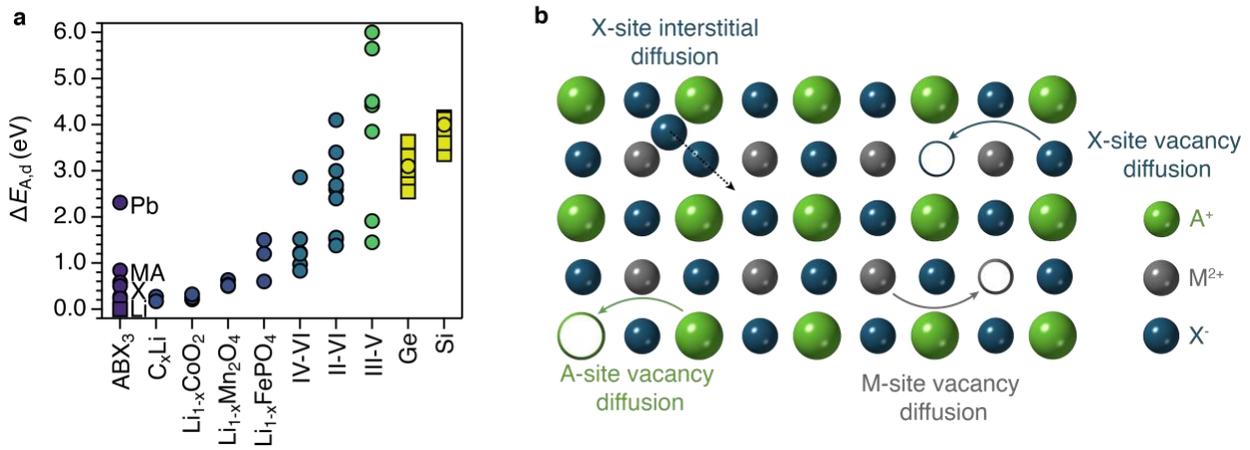

**Figure 4.** (a) Activation energy of diffusion ($E_{A,d}$) comparison of MHPs to conventional semiconductors and Li-ion battery electrodes. Circles denote intrinsic diffusion and squares denote extrinsic diffusion. (b) Illustration showing the common ion transport mechanisms in MHPs.

Ion transport in MHPs occurs through defects due to the charge imbalance introduced by the defect. The weak cohesive forces in the MHP lattice lead to high defect concentrations that can even be higher than carrier concentrations.[90] Each of the A-, B- and X-sites are able to migrate through vacancy-mediated ionic diffusion while X-sites are also able to migrate interstitially (Figure 4b).[83, 84] The corner sharing octahedral network of the MHP lattice gives each ion/vacancy eight nearest neighbors, making vacancy-mediated ionic diffusion especially prevalent. The high defect density inherent in MHPs, many nearest neighbors over short distances, and weak cohesive forces in the lattice result in low energy input for ions migration.

The high Ionic diffusivity of MHPs enables remarkable radiation hardness, and their high Z elements and long carrier lifetimes make them promising candidates for next-generation radiation detectors. While ionizing radiation can displace nuclei, creating vacancies and trap states, MHPs "self-heal" as displaced ions diffuse back to lattice sites and restore electronic quality. As a result, MHPs have a damage threshold to proton irradiation nearly three orders of magnitude higher than crystalline Si, which holds promise for non-terrestrial PV.[91-92] $CsPbBr_3$ also achieved energy resolution of 1.4%,[93] rivaling state-of-the-art CZT, and may offer orders of magnitude reduction in cost.[94] Polycrystalline thick film detectors have also been blade coated directly on a-Si TFT backplanes for direct conversion x-ray imaging.[95] Key challenges remain: single crystals size is limited, electrode diffusion is problematic, high dark currents due to low bulk resistivity and self-doping the reduce signal to noise ratio, and the spatial resolution of imaging detectors can improve.

High ion transport in MHPs renders them excellent materials for energy storage applications such as Li/Na-ion batteries, photorechargeable batteries, and as solid-state electrolytes.[96] The MHP lattice can be reversibly doped by a variety of ions including $Li^+$ and $Na^+$ *via* electrochemical methods.[97] Li/Na-ion batteries have been fabrication from a variety of MHP compositions including organic, inorganic, 2D, 1D, and transition metals.[96] Notably, low-dimensional perovskites have achieved 646 mAh/g and 961 mAh/g capacities for Li-ion and Na-ion batteries, respectively, exceeding that of conventional graphite anodes (372 mAh/g).[98] Copper-based 2D MHP have also been used as Li-ion cathodes with 200 cycles, but their specific capacity was an order of magnitude lower than commercial $LiNiCoMnO_2$ cathodes.[99] MHP can also undergo color changes upon $Li^+$ intercalation, and $CsPbBr_3$ electrodes exhibit electrochromism upon $Li^+$ intercalation between orange and gray/black.[97]

Unlike conventional battery materials, MHP's act as a battery electrode while simultaneously using sunlight to charge itself as a photorechargeable battery. The $Cs_3Pb_2I_9$ defect perovskite demonstrated this effect with a 975 mAh/g capacity under illumination.[100] $MA(Pb/Sn)Cl_3$ has also been used as a solid-state electrolyte, allowing Li-ion transport while suppressing dendrite formation at the surface of Li metal electrodes (specific capacity 3860 mAh/g).[27] Supercapacitors have specific energies 1-2 orders of magnitude greater than conventional electrolytic capacitors (~0.02 Wh/kg) and find a niche where the required charge/discharge rates and cycling life exceed the performance of rechargeable batteries, with applications in regenerative breaking and power grid buffering. 3D MAPbBr$_3$ and 2D PEA$_2$PbBr$_4$ supercapacitors, with energy densities up to 9 Wh/kg, retained 98% and ~100%, respectively, of their initial performance after charge 1,000 cycles. The 3D material offered greater capacitance and energy density, while the improved diffusion kinetics of the 2D material offered greater power density.[101]

High ion transport combined with responsive optoelectronic properties have enabled emerging technologies. Electronic ratchets utilize spatially asymmetric potential distributions to convert nondirectional sources of energy into direct current. Recently, the first MHP electronic ratchet used a voltage stress to redistribute ions within a 2D MHP, converting electronic noise and unbiased square-wave potentials into current.[31] LEC devices induce a p-i-n junction within the MHP layer *via* ion accumulation at interfaces to enable high emissivity,[32] with advantages over LEDs including simple architecture, cost-efficient fabrication, and air stable electrodes. However, the best LEC $t_{1/2}$ lifetime is 6700 h at 100 cd/m$^2$ and degradation mechanisms require deeper understanding. Humidity sensors have also been developed that undergo photoluminescence and resistance changes caused by ion transport induced by exposure to humidity.[102-104] Control of ion migration through a heterojunction formed between MHP nanocrystals and single-walled carbon nanotubes enables optical switching and functions for neuromorphic computing.[34] ReRAM utilizes electrochemical metallization to form and rupture metallic filaments, resulting in low and high resistance states. Challenges for MHP-based ReRAM include short retention times (thousands of seconds), low endurance (a few thousand cycles), and insufficient on/off ratios ($< 10^{10}$), but these may be overcome. Contracting the lattice by moving from MAPbI$_3$ to MAPbCl$_3$ increased the extrapolated retention time from 1.6 to 28.3 years in a perovskite ReRAM device.[105] The electric field-induced ion migration effect has also been used to reconfigure the photoresponsivity of MHP devices over the range of 540-1270%, which enables the fabrication of adaptive machine vision systems with a maximum 263% enhancement of object recognition accuracy.[106]

*Summary and Future Outlook.* The dynamic structural flexibility inherent in metal halide perovskites result from the union of several unique properties whose combination is not observed in other semiconductors. The low formation energies, low energy difference between polymorphs, and high ion transport inherent in MHPs allow its lattice to easily form, break apart, and rearrange with little energy input. Though these features are often problematic for conventional applications, they should not be feared; they offer a unique opportunity to realize emerging applications that require dynamic properties such as switchable PV smart windows, electronic ratchets, solid-state electrolytes, light-emitting electrochemical cells, and artificial synapses for neuromorphic computing. This perspective provides guidance on how to exploit the unique properties found in MHPs and offers a fresh perspective on the properties of these materials without the PV-centric narrative usually found in the literature.

## ASSOCIATED CONTENT

$\Delta_f H°$, $\Delta_f S°$, $\Delta_f G°$, $\Delta U_L°$, $E_{A,d}$ values used for all plots and their references.

## AUTHOR INFORMATION

**Corresponding Author**

*E-mail: Lance.Wheeler@nrel.gov

**ORCID**

Bryan A. Rosales: 0000-0003-2488-7446

Joseph J. Berry: 0000-0003-3874-3582

Lance M. Wheeler: 0000-0002-1685-8242

**Twitter**

Bryan A. Rosales: @ Bryan_A_Rosales

Joseph J. Berry: @Perovdaddy

Lance M. Wheeler: @LanceMWheeler

**Notes**

The authors declare no competing financial interest.

## ACKNOWLEDGMENT

This study was authored by the National Renewable Energy Laboratory, operated by Alliance for Sustainable Energy, LLC, for the U.S. Department of Energy (DOE) under contract No. DEAC36-08GO28308. Funding was provided by the Building Technologies Office within the U.S.

Department of Energy Office of Energy Efficiency and Renewable Energy. The views expressed in the article do not necessarily represent the views of the DOE or the U.S. Government. The U.S. Government retains and the publisher, by accepting the article for publication, acknowledges that the U.S. Government retains a nonexclusive, paid-up, irrevocable, worldwide license to publish or reproduce the published form of this study, or allow others to do so, for U.S. Government purposes.